\documentclass[twocolumn,notitlepage,prb,superscriptaddress,longbibliography]{revtex4-2}
\usepackage[T1]{fontenc}
\usepackage[latin9]{inputenc}
\usepackage{color}
\usepackage{babel}
\usepackage{amssymb}
\usepackage{graphicx}
\DeclareGraphicsExtensions{.pdf}
\usepackage{amsmath,bm,color}
\usepackage{float}
\usepackage{epstopdf}
\usepackage{tabularx}
\usepackage{braket}
\usepackage{booktabs}
\usepackage{array}
\usepackage{blindtext}
\usepackage{natbib}
\usepackage[unicode=true,pdfusetitle,
 bookmarks=true,bookmarksnumbered=false,bookmarksopen=false,
 breaklinks=false,pdfborder={0 0 1},backref=false,colorlinks=false]
 {hyperref}

\makeatletter

\usepackage{babel}
\usepackage{babel}

\begin{document}

\title{Higher-order topology induced by altermagnetism in quasicrystals}

\author{Xingmin Huo}
\affiliation{School of Physics, Beihang University,
Beijing 100191, China}

\author{Xingchuan Zhu}
\affiliation{School of Intelligence Science and Technology, Nanjing University of Science and Technology, Nanjing 210094, China}

\author{Xingyu Ma}
\affiliation{School of Mathematics and Physics, University of Science and Technology Beijing, Beijing 100083, China}

\author{Shiping Feng}
\affiliation{Faculty of Arts and Sciences, Beijing Normal University, Zhuhai 519087, China, and School of Physics and Astronomy,  Beijing Normal University, Beijing 100875, China}

\author{Huaiming Guo}
\email{hmguo@buaa.edu.cn}
\affiliation{School of Physics, Beihang University,
Beijing 100191, China}

\begin{abstract}
Altermagnetism (AM), a novel magnetic phase featuring zero net magnetization and momentum-dependent spin splitting, provides a compelling platform for exploring exotic quantum phenomena. While recent studies have extended the concept of AM to quasicrystals, the momentum-dependent spin splitting in the absence of translational symmetry remains inadequately understood. Here, we investigate the emergence of higher-order topological phases (HOTPs) in Thue-Morse and Ammann-Beenker quasicrystals, driven by the coupling between $d$-wave quasicrystal AM orders and a two-dimensional topological insulator. We reveal that the $d_{x^2-y^2}$-wave AM exhibits a ferromagnetism-like aspect in opening edge gaps and inducing zero-energy higher-order topological corner states. In contrast, conventional antiferromagnetic and $d_{xy}$-wave AM configurations are unable to induce the higher-order topological gaps. These findings highlight the pivotal role of altermagnetic symmetry and its alignment with lattice corners in the realization of HOTPs and establish that the emergence of HOTPs can serve as a signature for identifying altermagnetic orders in aperiodic systems.
\end{abstract}

\date{\today}

\maketitle
\section{Introduction}

Altermagnetism (AM) has recently attracted considerable attention as a distinct form of collinear magnetic order beyond the conventional framework of ferromagnetism and antiferromagnetism~\cite{PhysRevX.12.031042,PhysRevX.12.040501,AdvFunctMater.2409327,jungwirth2024altermagnetsbeyondnodalmagneticallyordered,magnetism5030017,npjSpintronics.3.1,Song2025,JUNGWIRTH2025100162,Jungwirth2026,liu2026altermagnetismsuperconductivityshorthistorical,rimmler2025noncollinear}. It combines a fully compensated magnetic structure with pronounced momentum-dependent spin splitting, even in the absence of relativistic spin-orbit coupling. This spin splitting becomes possible when the symmetries protecting spin degeneracy in conventional collinear antiferromagnets, such as combined $\mathcal{PT}$ symmetry, are absent~\cite{PhysRevX.14.031037,PhysRevX.12.021016,PhysRevX.14.031038,PhysRevX.14.031039,PhysRevLett.132.176702}. Spectroscopic, transport, and magnonic studies of materials such as MnTe, CrSb, KV$_2$Se$_2$O, and Rb$_{1-\delta}$V$_2$Te$_2$O have provided growing evidence for AM spin splitting and related responses~\cite{Krempasky2024,Nature2024Altermagnetism,PhysRevLett.132.036702,PhysRevLett.133.156702,Reimers2024,PhysRevLett.133.206401,Yang2025,Biniskos2025,zhou2025manipulation,Jiang2025,Zhang2025,7yhz-jptc,Zeng2024,Nature.626.523,lin2024observationgiantspinsplitting,SciAdv.10.eadj4883}. The combination of magnetic compensation, potentially ultrafast dynamics, and efficient electrical control of spin transport makes AM a promising platform for future spintronic applications~\cite{doi:10.1126/sciadv.adn0479,PhysRevLett.133.056401}.

AM integrated into topological insulators (TIs) can induce higher-order topological phases (HOTPs), providing a distinct pathway for manipulating topological corner states (TCSs)~\cite{PhysRevB.109.L201109,PhysRevB.109.245306,xbzb-jgk9,Wang2026,PhysRevB.111.155303,doi:10.1021/acs.nanolett.5c03191,9wcm-pmr2,g1wx-v888}. The characteristic momentum-dependent exchange field of AM generates Dirac masses that vary among the boundaries of a TI. When adjacent edges acquire masses of opposite sign, topological domain walls form at the corners, trapping localized TCSs. Crucially, the sign pattern of these boundary masses is determined by the orientation of the N\'{e}el vector, enabling the spatial repositioning of corner states via rotating the magnetic order. Compared to conventional ferromagnetic exchange fields, AM facilitates analogous higher-order topological responses while maintaining zero net magnetization. This mechanism has been extended to topological superconductors, giving rise to higher-order topological superconducting states that host localized Majorana zero modes~\cite{PhysRevB.109.224502,g17w-xs73,PhysRevB.108.184505,PhysRevLett.133.106601,PhysRevB.108.205410}. These developments establish AM as a promising route to generating and controlling higher-order topological states.

Recent studies have shown that AM order can emerge even in the absence of periodic crystal symmetry~\cite{zqnk-153f,shao2025quasicrystals}.
Quasicrystals are characterized by long-range quasiperiodic order without translational symmetry. Their intrinsic rotational symmetry makes them a natural setting for exploring AM beyond periodic crystals. Octagonal Ammann--Beenker and dodecagonal Stampfli quasicrystals have been predicted to host $g$-wave and $i$-wave AM protected by $C_8\mathcal{T}$ and $C_{12}\mathcal{T}$ symmetries, respectively~\cite{src1-b2xt}. Hyperspace projection has also revealed $g$-wave and $h$-wave AM in decorated Ammann--Beenker and Penrose quasicrystalline lattices, respectively~\cite{hx9n-v1yg}. Symmetry analysis combined with sign-problem-free projector quantum Monte Carlo simulations further indicates that bipartite quasicrystals with N\'{e}el order can realize antiferromagnetic, altermagnetic, or ferromagnetic phases depending on the symmetry relation between their sublattices~\cite{ji2025neelorderedmagneticphases}.  

While quasicrystals are expected to host a broad range of AM phases, characterizing AMs in quasicrystals is less straightforward due to the absence of translational symmetry. Without a conventional Brillouin zone or well-defined Bloch momentum, its characteristic momentum-dependent spin splitting cannot be directly resolved from conventional band dispersions and is instead inferred from spin-resolved spectral functions by projecting the eigenstates onto plane-wave states with definite momentum. 
Naturally, higher-order topology provides a compelling real-space alternative, as its signatures are natively encoded in boundary mass gaps and the resulting localized TCSs~\cite{PhysRevLett.124.036803,PhysRevB.104.245302,PhysRevB.102.241102,PhysRevB.108.195306,huang2021generic,PhysRevB.111.075306,PhysRevResearch.2.033071,PhysRevB.109.134205,c52f-31my}.  An AM exchange field can generate a symmetry-dependent pattern of boundary masses, whereas a corresponding conventional antiferromagnetic order may leave the helical edge states gapless. Whether this distinction survives in quasicrystals and can serve as a real-space feature of AM remains an interesting question.

In this paper, we systematically analyze the ferromagnetic-like characteristics of AM in driving higher-order topology and investigate the higher-order topological response of altermagnets in quasicrystals. We consider Thue-Morse and Ammann-Beenker quasicrystals coupled to a two-dimensional first-order TI, comparing the effects of ferromagnetic, $d_{x^2-y^2}$-wave AM, and $d_{xy}$-wave AM orders. By analyzing spin-resolved spectral functions, edge spectra, and the real-space distribution of corner states, we demonstrate that $d_{x^2-y^2}$-wave AM can induce higher-order topological edge gaps and bind TCSs. This behavior closely parallels that of a ferromagnetic exchange field, despite the altermagnet's vanishing net magnetization. In contrast, the $d_{xy}$-wave altermagnet is unable to generate higher-order corner modes. Our results not only establish quasicrystalline altermagnets as a controllable platform for engineering higher-order topological states but also demonstrate that the emergence of TCSs serves as a signature of AM phases in quasicrystals.

The paper is structured as follows: Section II analyzes the role of AM in driving higher-order topology on periodic square lattice. Sections III and IV examine AM-induced higher-order topological states in Thue-Morse and Ammann-Beenker quasicrystals, respectively. Section V presents the conclusions.

\section{Ferromagnetic aspect of AM in inducing higher-order topology}

\begin{figure*}[hbpt]
  \includegraphics[width=16cm]{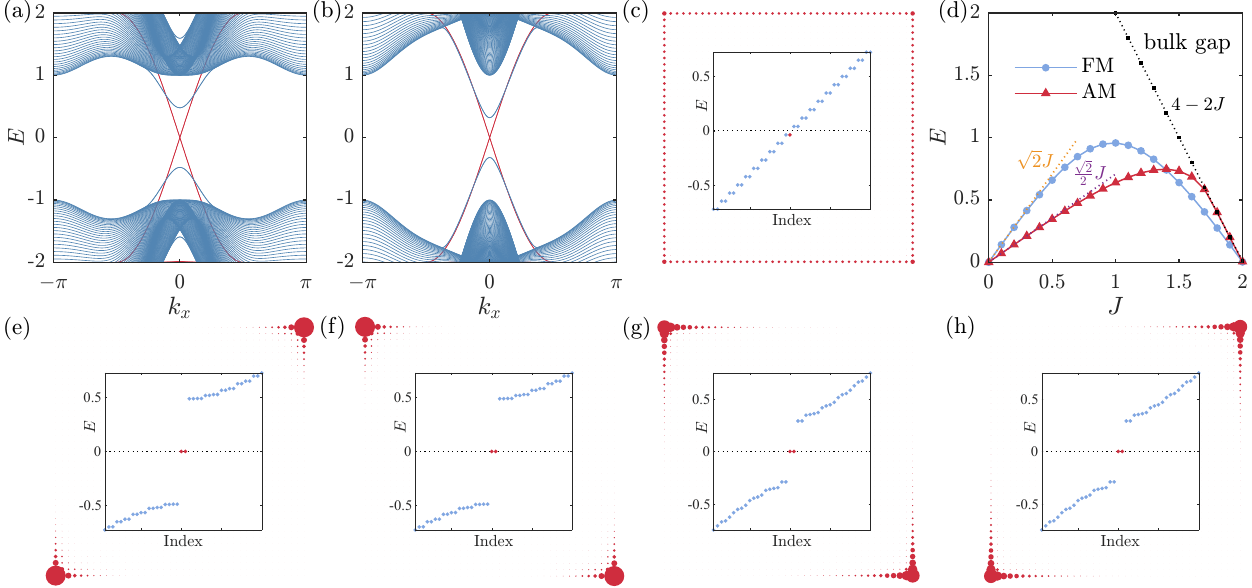}
  \caption{HOTPs induced by FM and $d_{x^2-y^2}$-wave AM. Energy spectra on a TI strip incorporating (a) FM and (b) AM orders. The red and blue curves represent $J$ = 0 and $J$ = 1, respectively. (c) Distribution of the topological edge states at $J$=0 for the eigenstate near the Fermi level (see inset). (d) Evolution of the edge-state gaps induced by the FM and AM orders as a function of the exchanging strength $J$. (e) and (f) Spatial distributions of the corner states in the HOTP induced by FM order, with magnetic moments along the [1$\bar{1}$] and the [11] directions, respectively. (g) and (h) Corresponding distributions for the $d_{x^2-y^2}-$wave AM. The parameters are set as $m_0 =2,t=1$ and $A=1$.}\label{fig1}
\end{figure*}

AM has recently been identified as a mechanism capable of inducing HOTPs in TIs. This phenomenon arises from the characteristic momentum-dependent spin splitting of altermagnets, which manifests certain ferromagnetic-like features despite a vanishing net magnetization. To better elucidate this effect, we first perform a systematic analysis of how AM drives higher-order topology in periodic systems based on the two-dimensional Bernevig-Hughes-Zhang (BHZ) model on a square lattice. Such an analysis is intended to establish the theoretical foundation and provids the physical intuition necessary to understand AM-induced HOTPs in quasicrystals, where the underlying mechanisms remain similar. The Hamiltonian in the basis
$\Psi_k =\left(c_{s,\uparrow}, c_{p,\uparrow}, c_{s,\downarrow}, c_{p,\downarrow}\right)^T$ is given by
\begin{align}
H_0(\boldsymbol{k})=M(\boldsymbol{k})\sigma_{z}+2A\sin k_{x}s_{y}\sigma_{x}+2A\sin k_{y}s_{x}\sigma_{x},
\label{eq1}
\end{align}
where $M(\boldsymbol{k}) = \left( m_0 - 2t \cos k_x - 2t \cos k_y \right)$, and $s_i \ \text{and} \ \sigma_i \quad (i=x,y,z)$ are Pauli matrices acting on the orbital and spin degrees of freedom, respectively.
The topological properties of this Hamiltonian are characterized by a $\mathbb{Z}_2$ topological invariant, which is determined by the parity eigenvalues at the four time-reversal-invariant momenta~\cite{PhysRevB.76.045302}. For $0 < |m_0| < 4t$, the system resides in a TI phase, hosting gapless helical edge states that traverse the bulk energy gap.

To systematically investigate the effects of various magnetism, we augment the BHZ model with different magnetic orders, including a ferromagnetic (FM) phase and both $d_{x^2-y^2}$- and $d_{xy}$-wave AM configurations~\cite{PhysRevB.109.L201109,ganguli2025tunable,PhysRevB.111.125420}. The corresponding exchange terms in momentum space are given by:
\begin{subequations}
\begin{align}
&H_{\text{FM}}(\boldsymbol{k}) = J \, \boldsymbol{s} \cdot \hat{\boldsymbol{n}} , \label{eq2a} \\
&H_{d_{x^2-y^2}}(\boldsymbol{k}) = \frac{J}{2} \left( \cos k_{x} - \cos k_{y} \right) \, \boldsymbol{s} \cdot \hat{\boldsymbol{n}} , \label{eq2b} \\
&H_{d_{xy}}(\boldsymbol{k}) = J \, \sin k_{x} \sin k_{y} \, \boldsymbol{s} \cdot \hat{\boldsymbol{n}}, \label{eq2c}
\end{align}
\end{subequations}
where $\hat{\mathbf{n}} = (\sin\theta\cos\phi, \sin\theta\sin\phi, \cos\theta)$ denotes the direction of the N\'{e}el vector, and $J$ represents the strength of the exchange interaction. 

In the ferromagnetic case, when the magnetization direction $\hat{\boldsymbol{n}}$ is aligned along the in-plane $[1\bar{1}]$ direction (i.e., $\theta=\pi/2$ and $\phi=-\pi/4$), a gap opens in the topological edge states at $k_x=0$. Correspondingly, the energy spectrum on a finite-sized square sample exhibits two zero-energy mid-gap states. Furthermore, the wavefunctions of these zero-energy states are highly localized at the bottom-left and top-right corners of the sample [see Fig.~\ref{fig1}(e)]. The emergence of TCSs signifies  the realization of an HOTP induced by ferromagnetic order in the TI. Notably, when the N\'{e}el vector $\hat{\boldsymbol{n}}$ is rotated to the $[11]$ direction, the TCSs shift to the top-left and bottom-right corners, as shown in Fig.~\ref{fig1}(f). Consequently, tuning the orientation of the N\'{e}el vector provides an effective means to manipulate the spatial localization of these corner states. Conversely, the spatial distribution of the TCSs can be utilized to determine the underlying direction of the N\'{e}el vector, establishing a direct correspondence between magnetic order and topological signatures.

Replacing the FM order with a $d_{x^2-y^2}$-wave AM order leads to a qualitatively similar effect on the TI edge states. As shown in Fig.~\ref{fig1}(b), the AM order gaps out the helical edge modes. In the small $J$ regime, however, the induced gap is approximately one-half of that generated by the FM order for the same exchange coupling strength $J$. The energy spectrum of a finite-sized square sample, displayed in the inset of Fig.~\ref{fig1}(g), exhibits two in-gap states pinned at zero energy. Examination of their spatial distributions confirms that these states are localized at the sample corners, indicating the emergence of TCSs. Therefore, similar to the FM case, the $d_{x^2-y^2}$-wave AM order drives the system into an HOTP.
The induced TCSs also exhibit a pronounced orientation dependence. As illustrated in Fig.~\ref{fig1}(g), when the  N\'{e}el vector $\hat{\mathbf{n}}$ is aligned along the $[1\bar{1}]$ direction, the corner states are predominantly localized at the top-left and bottom-right corners of the square sample. Upon rotating $\hat{\mathbf{n}}$ toward the $[11]$ direction, the TCSs migrate to the opposite diagonal corners, as shown in Fig.~\ref{fig1}(h). This behavior closely parallels that observed in the FM case, further demonstrating the common higher-order topological origin arising from the spin-split bands.
It is noted that the spatial symmetry of the altermagnetic order plays a crucial role in realizing the HOTP. In contrast to the $d_{x^2-y^2}$-wave AM order, the $d_{xy}$-wave AM order fails to gap out the helical edge states of the TI and therefore cannot drive the system into an HOTP.

To elucidate the intrinsic relationship between the orientation of the N\'{e}el vector and the locations of the corner states, we perform an analytical study based on the low-energy effective theory. Since the bulk gap opens at the $\Gamma=(0,0)$ point, we expand the Hamiltonian around this momentum, yielding

\begin{align}
{\cal H}_{0}(k)= &(m_0-4t+t k_x^2+t k_y^2)\sigma_z 
+ \label{eq3}\\
&2Ak_xs_y\sigma_x +2Ak_ys_x\sigma_x.
\nonumber
\end{align}
The low-energy Hamiltonians for the FM, $d_{x^2-y^2}$-wave and $d_{xy}$-wave AMs are expanded as
\begin{subequations}
\begin{align}
&{\cal H}_{\text{FM}}(\boldsymbol{k}) = J \, \boldsymbol{s} \cdot \hat{\boldsymbol{n}} , \label{eq4a} \\
&{\cal H}_{d_{x^2-y^2}}(\boldsymbol{k}) = -\frac{J}{4} \left(  k_{x}^2 - k_{y}^2 \right) \, \boldsymbol{s} \cdot \hat{\boldsymbol{n}} , \label{eq4b} \\
&{\cal H}_{d_{xy}}(\boldsymbol{k}) = J \,  k_{x}  k_{y} \, \boldsymbol{s} \cdot \hat{\boldsymbol{n}} . \label{eq4c}
\end{align}
\end{subequations}

We first analyze the FM case, described by the total Hamiltonian ${\cal H}(\boldsymbol{k})={\cal H}_0(\boldsymbol{k})+{\cal H}_{FM}(\boldsymbol{k})$. To derive the low-energy theory for the edge states, we solve the above Hamiltonian in the semi-infinite geometry $y \geq 0$. Replacing $k_y$ by $-\mathrm{i}\partial_{y}$,  the Hamiltonian takes the form ${\cal H}\left(k_{x},-\mathrm{i}\partial_{y}\right)={\cal {H}}_{0}\left(k_{x},-\mathrm{i}\partial_{y}\right)+{\cal H}_{FM}\left(k_{x},-\mathrm{i}\partial_{y}\right)$. In the small $J$ limit, ${\cal H}_{FM}$ is treated as a perturbation. We first solve the unperturbed eigenvalue equation ${\cal H}_{0}\left(-\mathrm{i}\partial_{y}\right)\psi(y)=E\psi(y)$, subject to the boundary conditions $\psi|_{y=0}=\psi|_{y=+\infty}=0$. The two zero-energy solutions are given by $\psi_{\alpha}(y) = C \sin(\gamma_{1} y) e^{-\gamma_{2} y} \chi_{\alpha}$, where $\gamma_{1}=\sqrt{(4t-m_0) t-A^{2}}/t, \quad \gamma_{2}=\frac{A}{t}$, and $\chi_1 =\left(\begin{array}{l} i, 0, 0, 1 \end{array}\right)^T/\sqrt{2}$ and $\chi_2 =\left(\begin{array}{l} 0, -i, 1, 0 \end{array}\right)^T/\sqrt{2}$. The normalization constant is $C=\sqrt{4\gamma_{2}\left(\gamma_{1}^{2}+\gamma_{2}^{2}\right)/\gamma_{1}^{2}}$. Projecting the perturbation term \({\cal H}_{FM}\) onto the zero-energy edge-state basis \((\psi_1,\psi_2)\), we obtain the effective boundary Hamiltonian: 
\begin{align}
{\cal H}_{\rm eff}(k_x)=-2A k_x s_z+J\cos\phi\,s_y, 
\end{align}
which is precisely the one-dimensional massive Dirac equation.

Generally, the effective Hamiltonians for the four boundaries of a finite-sized square sample can be expressed as
\begin{align}
{\cal H}_{\rm eff}^{(\ell)}\left(k_x\right)= -2A k_x \boldsymbol{s}_{z}+m(\ell) \boldsymbol{s}_{y},
\label{heff}
\end{align}
where the mass term $m(\ell)$ varies among different boundaries. For the four boundaries $\ell=1,2,3,4$, ordered counter-clockwise starting from the bottom one, $m(\ell)$ takes the values $J{\rm cos}\phi$, $-J{\rm sin}\phi$, $-J{\rm cos}\phi$
and $J{\rm sin}\phi$, respectively. For $\phi=-\pi/4$, the mass term $m(\ell)$ reverses sign at the bottom-left and top-right corners. According to Jackiw-Rebbi theory, the mass domain walls formed thereby host zero-energy bound states. When $\phi$ is changed to $\pi/4$, the mass domain walls shift to the complementary corners.  These results are consistent with the spatial distributions presented in Fig.~\ref{fig1}(e) and (f).

For the case with the $d_{x^2-y^2}$-wave altermagnetic phase, the effective Hamiltonian retains the form of Eq.~(\ref{heff}), while the boundary mass terms $m(\ell)$ are given by $J_{\rm AM}{\rm cos}\phi$, $J_{\rm AM}{\rm sin}\phi$, $-J_{\rm AM}{\rm cos}\phi$ and $-J_{\rm AM}{\rm sin}\phi$ for the four edges in sequence, where $J_{\rm AM}=J(4t-m_0)/(4t)$. For a given value of $\phi$, the corner states appear at corners complementary to those in the FM phase. In contrast, for the $d_{xy}$-wave altermagnetic phase, all boundary mass terms vanish, $m(\ell)=0$, implying that the topological edge states remain gapless, consequently, no HOTP can be induced.

From the effective edge Hamiltonian in Eq.~(\ref{heff}), the edge gap is given by $\Delta^{(\ell)}=|m(\ell)|$. For the FM phase, the gap reads $\Delta_{\rm FM}=2J\cos\phi$, whereas for the altermagnetic phase it becomes $\Delta_{\rm AM}=J(4t-m_0)\cos\phi/(2t)$. As shown in Fig.~1(d), in the small-$J$ regime, the analytical predictions are in good agreement with the results obtained from numerical diagonalization. For the representative parameters $t=1$ and $m_0=2$, the edge gap in the FM phase is twice that in the altermagnetic phase.

The bulk gap follows the linear relation $\Delta_{\rm bulk}=2m_0-2J$, as shown in Fig.~\ref{fig1}(d). While the effective edge theory predicts a linear increase of the edge-state gap with $J$, this behavior is limited by the shrinking bulk gap, which bounds the energy window available for the edge states. Consequently, the edge-state gap gradually departs from the linear trend at larger $J$ and eventually decreases together with the bulk gap. When $J=m_0$, the bulk gap closes, causing the edge states to merge into the bulk continuum and disappear.

\begin{figure}[hbpt]
  \includegraphics[width=8.5cm]{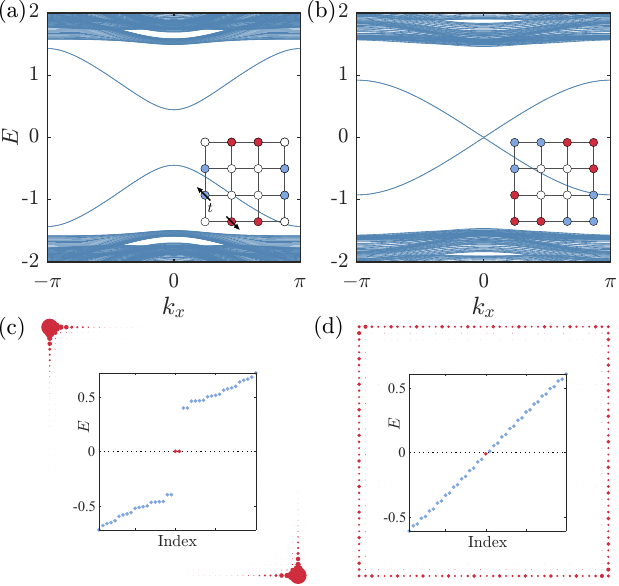}
  \caption{Energy spectra on a square-lattice TI strip incorporating AM orders with (a) $d_{x^2-y^2}$-wave and (b) $d_{xy}$-wave symmetries. 
  The insets of (a) and (b) illustrate the corresponding altermagnetic configurations. (c) Real-space distributions of the corner states in the HOTP induced by the AM order in (a). (d) Spatial distribution of the topological edge states in the presence of the AM order in (b).  The N\'{e}el vector in both (c) and (d) is oriented along the $[1\bar{1}]$ direction.  The model parameters are $m_0=2$, $t=1$, $A=1$, and $J=1$.}\label{fig2}
\end{figure}

\begin{figure*}[hbpt]
  \includegraphics[width=13cm]{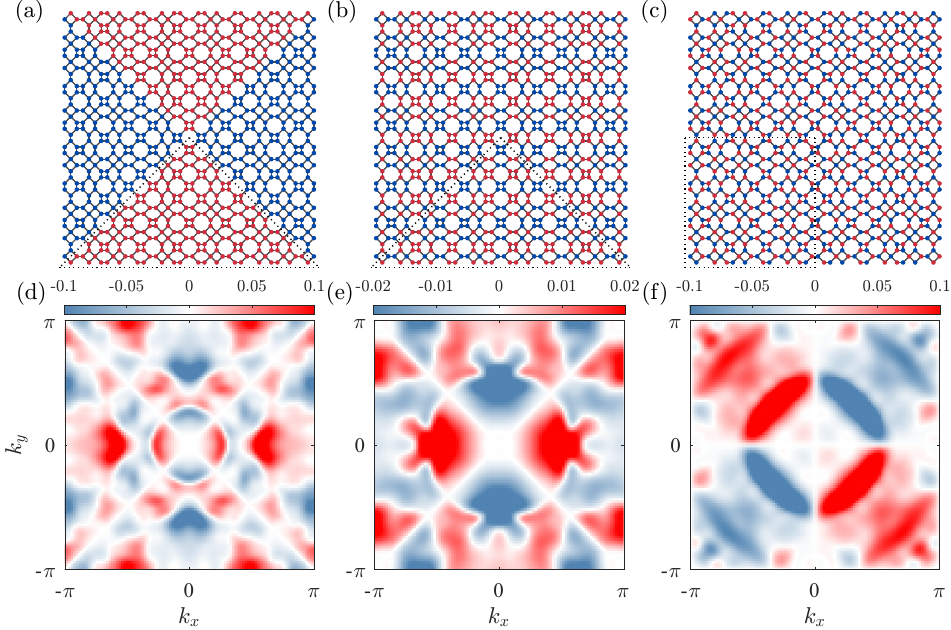}
  \caption{Altermagnetic configurations on Thue-Morse quasicrystals with: (a) and (b) $d_{x^2-y^2}$-wave symmetry; (c) $d_{xy}$-wave symmetry. The dashed triangle indicates the seed cluster used to generate the configurations via the $C_{4z}\mathcal{T}$ operation. (d)-(f) Momentum-space spin-resolved spectral density difference, $A_\uparrow(\boldsymbol{k}, \omega) - A_\downarrow(\boldsymbol{k}, \omega)$, at fixed energy levels: $\omega=-1.3$ for (d), and $\omega=0.3$ for (e) and (f). The parameters are set as $t=1$ and $J=1$.}\label{fig3}
\end{figure*}

\begin{figure}[hbpt]
  \includegraphics[width=8cm]{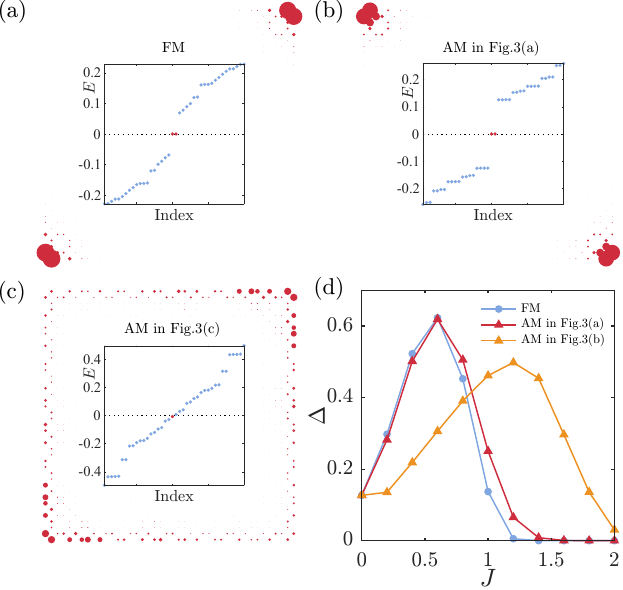}
  \caption{Spatial distributions of the corner modes in HOTPs on Thue-Morse quasicrystal induced by (a) the FM order and (b) the $d_{x^2-y^2}$-wave AM in Fig.~\ref{fig3}(a). The insets in (a) and (b) show the low-energy spectrum, revealing the presence of two zero-energy corner modes. (c) Real-space distribution of the topological edge states in the presence of $d_{xy}$-wave AM. The plotted state corresponds to the eigenstate near the Fermi level indicated in the inset. (d) Evolution of the higher-order-topology gap as a function of the exchange coupling $J$ for the FM and the two $d_{x^2-y^2}$-wave AM configurations. The calculations are performed on a lattice of $880$ sites with the N\'{e}el vector oriented along the $[1\bar{1}]$ direction. Parameters are fixed at $t=1, m_0=2$, and $A=1$. In (a)--(c), the exchange coupling is set to $J=1$.}
  \label{fig4}
\end{figure}

The above analysis is based on the effective description of AM, in which the underlying magnetic structure is not explicitly encoded. To make the magnetic configuration manifest, we now consider a lattice Hamiltonian that explicitly incorporates the altermagnetic order [see the insets of Fig.~\ref{fig2} for the $d_{x^2-y^2}$-wave and $d_{xy}$-wave altermagnetic configurations]. The real-space Hamiltonian describing the TI coupled to the altermagnetic order can be generally written as
\begin{align}
H= &-\sum_{\langle i j\rangle} c_{i}^{\dagger}\left[t \sigma_{z}+i A\left(s_{y}\sigma_{x}\cos\varphi_{i j}+s_{x}\sigma_{x}\sin\varphi_{i j}\right)\right]c_{j} \nonumber \\ 
& +m_0\sum_{i} c_{i}^{\dagger} \sigma_{z} c_{i} +J \sum_{i} c_{i}^{\dagger} \boldsymbol{s} \cdot \hat{\boldsymbol{n}} c_{i}, \label{eq7}
\end{align}
where the notation $\langle ij \rangle$ labels nearest-neighbor pairs with a hopping amplitude $t$, and $\varphi_{ij}$ represents the azimuthal angle of the bond connecting sites $i$ and $j$ with respect to the horizontal direction. 

Figure~\ref{fig2}(c) shows the energy spectrum for the $d_{x^2-y^2}$-wave altermagnetic case with the direction of the magnetic moment $\hat{\mathbf{n}}=[1\bar{1}]$. The altermagnetic order opens a gap in the edge spectrum, leading to the emergence of zero-energy corner states localized at the top-left and bottom-right corners of a finite square sample, thereby realizing an HOTP. Upon rotating $\hat{n}$ by $90^\circ$, the corner states relocate to the other pair of diagonally opposite corners. In contrast, the $d_{xy}$-wave altermagnetic order leaves the edge spectrum gapless, with the corresponding states remaining extended along the sample boundaries. These results are fully consistent with the preceding analysis, thus further validating the role of AM in inducing an HOTP.

\section{Higher-order topology by AMs on Thue-Morse quasicrystal}

We extend our analysis to quasicrystalline systems, beginning with the Thue-Morse quasicrystal. We construct several AM phases characterized by $d_{x^2-y^2}$- or $d_{xy}$-wave symmetries, where the magnetic sublattices are related by a fourfold rotation combined with time reversal ($C_{4z} \mathcal{T}$). In practice, we define a 'seed cluster' encompassing one-quarter of the square-shaped lattice and prescribe its magnetic moments~\cite{PhysRevLett.134.166701}. The global magnetic configuration is then generated by applying successive $C_{4z}\mathcal{T}$ operations to this seed. While this construction inherently ensures zero net magnetization, the simultaneous breaking of $P\mathcal{T}$ symmetry will lead to $d$-wave symmetric spin splitting, thereby realizing $d$-wave AM in the Thue-Morse quasicrystal.

Specifically, the region $y < -|x|$ is selected as the seed cluster to construct two distinct $d_{x^2-y^2}$-wave altermagnets. While all sites in Fig.~\ref{fig3}(a) are initialized in the spin-up state, the lattice in Fig.~\ref{fig3}(b) is configured with parallel spin orientations for sites connected by shorter bonds and antiparallel orientations for those connected by longer bonds. Both magnetic configurations satisfy the combined $C_{4z}\mathcal{T}$ symmetry but break $P\mathcal{T}$ symmetry.
Fig.~\ref{fig3}(c) displays a $d_{xy}$-wave altermagnetic configuration, generated by applying the $C_{4z}\mathcal{T}$ operation to the region $x<0, y<0$ (initially set to an AFM order). Notably, the resulting configuration differs from a standard AFM order, as the bonds crossing the diagonals exhibit ferromagnetic alignment.

To investigate the properties of the constructed AMs, we employ a spin-fermion tight-binding model in which the static local moments are determined by the altermagnetic configurations described above:
\begin{align}\label{eq1}
H=&-t \sum_{\langle i j\rangle, \alpha} c_{i, \alpha}^{\dagger} c_{j, \alpha}+J \sum_{i, \alpha, \beta} \boldsymbol{S}_i \cdot c_{i, \alpha}^{\dagger} \boldsymbol{\sigma}_{\alpha \beta} c_{i, \beta} \nonumber \\ 
&-\mu \sum_{i, \alpha} c_{i, \alpha}^{\dagger} c_{i, \alpha},
\end{align}
where $c_{i,\alpha}^{\dagger}$ and $c_{i,\alpha}$ are fermionic creation and annihilation operators with spin $\alpha\in\{\uparrow,\downarrow\}$ at site $i$. The parameter $J$ denotes the on-site exchange coupling between the static spin $\mathbf{S}_i$ and the itinerant electron spin $\boldsymbol{\sigma}$. The chemical potential $\mu$ controls the electron filling of the system.

To demonstrate the spin-dependent electronic structure in momentum space, the spin-resolved spectral function is typically employed for quasicrystals, defined as:
\begin{align}\label{eq8}
\mathcal{A}_{s}(\omega, \mathbf{k}) = \frac{\eta}{\pi} \sum_{n} \frac{|\langle n | \mathbf{k}, s \rangle|^{2}}{(\omega - E_{n})^{2} + \eta^{2}},
\end{align}
where $E_n$ and $|n\rangle$ denote the eigenvalues and eigenstates of the Hamiltonian, respectively, and $\eta$ is a small Lorentzian broadening parameter. Notably, since quasicrystals lack periodic translational symmetry and a formal Brillouin zone, the plane-wave state $|\mathbf{k}, s\rangle$ is utilized here as a projection basis to resolve the electronic states in momentum space. This basis state is defined as $|\mathbf{k}, s\rangle = \frac{1}{\sqrt{N}} \sum_j e^{i\mathbf{k} \cdot \mathbf{R}_j} |\mathbf{R}_j, s\rangle$, where $\mathbf{R}_j$ represents the position of the $j$-th site and $N$ is the total number of lattice sites.
Figures~\ref{fig3}(d)--(f) display the spin-resolved spectral density difference, $\delta A(\mathbf{k}) = A_\uparrow(\mathbf{k}) - A_\downarrow(\mathbf{k})$, corresponding to the magnetic configurations in Figs.~\ref{fig3}(a)--(c), respectively. The spectral differences in Figs.~\ref{fig3}(d) and (e) exhibit $C_{4z}\mathcal{T}$ symmetry, with values vanishing along the diagonal directions, which is a characteristic signature of $d_{x^2-y^2}$-wave altermagnet. In contrast, the spectrum in Fig.~\ref{fig3}(f) vanishes along the principal axes ($k_x$ and $k_y$), consistent with the symmetry of a $d_{xy}$-wave altermagnet.

\begin{figure}[hbpt]
  \includegraphics[width=8cm]{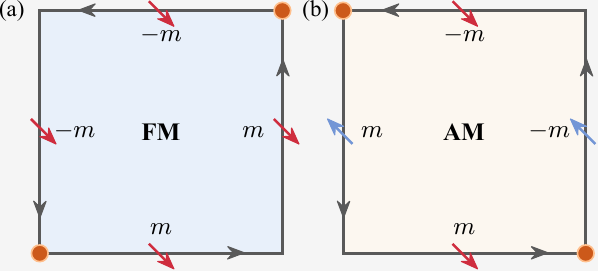}
  \caption{Schematic illustration of the boundary mass patterns in the effective edge Hamiltonian for the FM (a) and AM (b) cases.}
  \label{fig5}
\end{figure}

\begin{figure*}[hbpt]
  \includegraphics[width=13.cm]{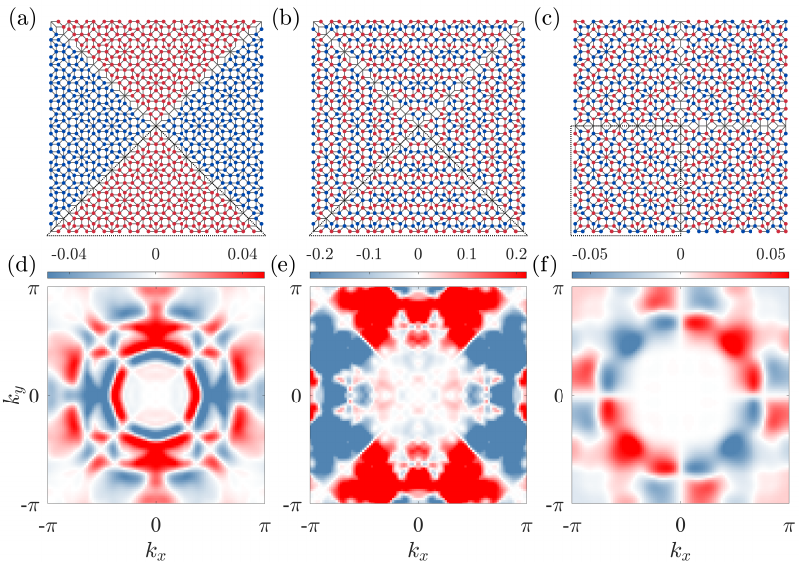}
  \caption{Altermagnetic configurations on Ammann-Beenker quasicrystals with: (a) and (b) $d_{x^2-y^2}$-wave symmetry; (c) $d_{xy}$-wave symmetry. The dashed triangle indicates the seed cluster used to generate the configurations via the $C_{4z}\mathcal{T}$ operation. (d)-(f) Momentum-space spin-resolved spectral density difference, $A_\uparrow(\boldsymbol{k}, \omega) - A_\downarrow(\boldsymbol{k}, \omega)$, at fixed energy levels: $\omega=-1.8$ for (d), $\omega=0$ for (e), and $\omega=-0.7$ for (f). The parameters are set as $t=1$ and $J=1$.}
  \label{fig6}
\end{figure*}

\begin{figure}[hbpt]
  \includegraphics[width=8.0cm]{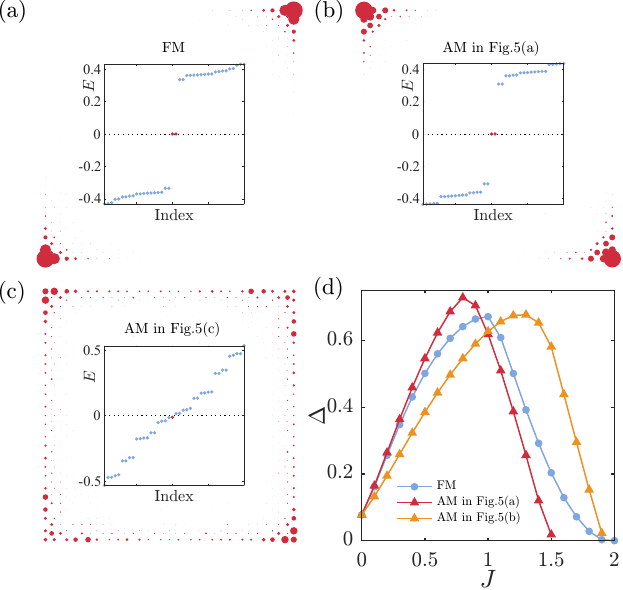}
  \caption{Spatial distributions of the corner modes in HOTPs on Ammann-Beenker quasicrystal induced by (a) the FM order and (b) the $d_{x^2-y^2}$-wave AM in Fig.~\ref{fig5}(a). The insets in (a) and (b) show the low-energy eigenspectra, revealing two zero-energy corner modes. (c) Spatial distribution of the topological edge states for the eigenstate near the Fermi level (see inset) in the presence of $d_{xy}$-wave AM. (d) Evolution of the energy gap as a function of the exchange coupling $J$ for the FM and the two $d_{x^2-y^2}$-wave AM configurations. The N\'{e}el vector is oriented along the $[1\bar{1}]$ direction. Parameters are fixed at $t=1, m_0=2$, and $A=1$. In (a)--(c), the exchange coupling is set to $J=1$.}
  \label{fig7}
\end{figure}

Next, we validate the constructed AMs on Thue-Morse quasicrystals by examining their capacity to induce HOTPs within a TI framework. We consider an open square geometry and couple the TI Hamiltonian in Eq.~\eqref{eq7} to the magnetic configurations, assuming the N\'{e}el vector is oriented along the $[1\bar{1}]$ direction. Initially, the FM order is considered as a benchmark. As shown in Fig.~\ref{fig4}(a), the FM order induces two in-gap states localized at the bottom-left and top-right corners, signaling the emergence of an HOTP. When the FM order is replaced by the $d_{x^2-y^2}$-wave AM orders depicted in Figs.~\ref{fig3}(a) and (b), an HOTP is similarly realized. 
Consistent with the analysis in Sec.~II, the corner states in the $d_{x^2-y^2}$-wave AM cases shift to the opposite diagonal corners compared to the FM case, as illustrated in Fig.~\ref{fig4}(b). Crucially, the induction of these HOTPs provides robust indirect evidence supporting the existence of AM order in quasicrystals. In contrast, for the $d_{xy}$-wave AM, the edge spectrum remains gapless, precluding the formation of an HOTP. This absence of an HOTP aligns perfectly with the discussion in the previous section.
Finally, Figure~\ref{fig4}(d) presents the edge gap as a function of the exchange coupling $J$ for the FM and two $d_{x^2-y^2}$-wave AM cases. In all cases, the gap exhibits a similar non-monotonic behavior: it initially widens with increasing $J$, attains a maximum, and subsequently narrows. The primary distinction between the two AM orders lies in the peak position of the gap, whereas the evolution curve for the $d_{x^2-y^2}$-wave AM configurations in Fig.~\ref{fig3}(a) is nearly identical to that of the FM order.

The emergence of the HOTI can be understood within the framework of Jackiw-Rebbi theory, in which sign changes of the boundary mass generate domain-wall bound states at the corners. The $C_{4z}\mathcal{T}$ symmetry implies the presence of inversion symmetry ${\cal P}$. In addition, the system preserves a mirror symmetry ($\mathcal{M}_{1\overline{1}}=i (s_x-s_y)/\sqrt{2}$) combined with a reversal of the magnetic moment $\mathcal{M}'_{1\overline{1}}$.
 The pattern of boundary mass signs is protected by inversion symmetry $\mathcal{P}$ and the modified mirror symmetry $\mathcal{M}'_{1\overline{1}}$. Under inversion [$\mathcal{P}{\cal H}(k)\mathcal{P}^{-1}={\cal H}(-k)$], the effective Hamiltonian on the $\ell$-th edge containing ${\cal H}_0=vk_x\boldsymbol{s}_z$ and ${\cal H}_m=m\boldsymbol{s}_y$ transforms as
\begin{align}
\mathcal{P}({\cal H}_0+{\cal H}_m)\mathcal{P}^{-1}= -({\cal H}_0-{\cal H}_m),
\end{align}
indicating that the effective mass changes sign on the opposite edge. Furthermore, the mirror symmetry $\mathcal{M}'_{1\overline{1}}$ enforces opposite mass signs on adjacent edges, i.e., $\mathcal{M}'_{1\overline{1}}({\cal H}_0+{\cal H}_m)\mathcal{M'}_{1\overline{1}}^{-1}= {\cal H}_0-{\cal H}_m$. Altogether, these symmetry constraints enforce the boundary mass pattern shown in Fig.~\ref{fig5}(b), thereby guaranteeing the emergence of corner states.

The HOTP can generally be characterized by a quantized topological invariant. We calculate the quadrupole moment, which is defined as\cite{PhysRevB.110.L121301,PhysRevResearch.2.012067},
\begin{align}
Q_{xy}=\left[\frac{1}{2\pi}\operatorname{Im}\ln\det\left(U_o^\dagger \hat{D}U_o\right)-Q_0\right]\bmod 1,
\tag{3}
\end{align}
where \(U_o=(|\psi_1\rangle,|\psi_2\rangle,\ldots,|\psi_{N_{\mathrm{occ}}}\rangle)\) is a matrix composed of the occupied eigenstates of the real-space Hamiltonian, \(\hat{D}=\exp(2\pi i\hat{x}\hat{y}/L^2)\), with \(\hat{x}\) and \(\hat{y}\) denoting the coordinate operators, and \(Q_0\) representing the contribution from the background positive charge distribution. We find that the system exhibits a quantized quadrupole moment \(Q_{xy}=0.5\) in the presence of the corner states, providing further evidence for the underlying higher-order topology. 

\section{Higher-order topology by AMs in Ammann-Beenker quasicrystal}

To further demonstrate the generality of employing induced HOTPs as a diagnostic for AM in quasicrystals, we extend our study to the Ammann-Beenker (AB) quasicrystal, which possesses eightfold rotational symmetry. Using the same construction framework, we realize altermagnetic configurations with $d_{x^2-y^2}$- and $d_{xy}$-wave symmetries on the AB lattice. These configurations are found to break global $\mathcal{PT}$ symmetry while preserving the combined $C_{4z}\mathcal{T}$ symmetry, thus satisfying the symmetry requirements for AM. By calculating the spin-resolved spectral density difference, $\delta A(\boldsymbol{k},\omega)$, we can characterize the momentum-space spin splitting. For the configurations in Figs.~\ref{fig6}(a) and \ref{fig6}(b), the spectral density difference vanishes along the diagonals, reflecting $d_{x^2-y^2}$-wave symmetry. In contrast, the configuration in Fig.~\ref{fig6}(c) displays a characteristic $d_{xy}$-wave symmetry in momentum space [Fig.~\ref{fig6}(f)], with nodal lines aligned along the principal axes.

Next, we validate these altermagnetic phases by incorporating them into BHZ-type TIs to investigate the emergence of HOTPs. We first consider the FM configuration aligned along the $[1\bar{1}]$ direction as a benchmark. As illustrated in Fig.~\ref{fig7}(a), a pair of zero-energy corner states appears, localized at the bottom-left and top-right corners of the square sample. This indicates that the induction of HOTPs by FM order in TIs is robust and independent of the specific lattice geometry. We then replace the FM order with the $d_{x^2-y^2}$-wave AM [Fig.~\ref{fig7}(b)]. A similar higher-order topological response is observed under the same conditions, as evidenced by the emergence of in-gap corner states. As expected, these corner states appear at the opposite pair of corners relative to the FM case. Combined with the vanishing net magnetization, this FM-like higher-order topological response provides compelling evidence for the altermagnetic nature of the system.
For comparison, Fig.~\ref{fig7}(c) illustrates the case for the $d_{xy}$-wave AM order under identical parameters. Unlike the  $d_{x^2-y^2}$ case, the $d_{xy}$-wave AM is unable to gap the edge states, leaving the topological boundary modes intact. This underscores that the specific spatial symmetry of the AM order is essential for opening boundary gaps and realizing the HOTP. Finally, we systematically investigate the evolution of the higher-order topological gap as a function of $J$, as presented in Fig.~\ref{fig7}(d). It is observed that the gap evolution for both AM phases follows a trend similar to that of the FM phase.

The mechanism by which corner modes emerge from boundary mass sign changes also applies to the AM-induced HOTPs in the Ammann-Beenker quasicrystal. The corresponding higher-order topology can be characterized by a quantized quadrupole moment.


\section{Conclusions}
We have systematically investigated the induction of HOTPs by altermagnetic orders in both periodic and quasicrystalline TIs, proposing that the emergence of HOTPs can serve as a robust diagnostic signature for AM. By coupling $d$-wave AM configurations to a BHZ TI on square lattice, we demonstrate that $d_{x^2-y^2}$-wave AM effectively gaps the helical edge states and generates zero-energy corner states, a hallmark of higher-order topology. Through an analytical low-energy effective theory and Jackiw-Rebbi domain-wall analysis, we elucidate that this higher-order topological response originates from the sign-reversal of boundary mass terms. Notably, the spatial localization of these corner states is determined by the orientation of the N\'{e}el vector, providing an intuitive way to infer the underlying magnetic orientation directly from the corner-state positions.

We subsequently extend the study to quasicrystalline systems, specifically the Thue-Morse and Ammann-Beenker structures. By employing a construction scheme based on seed clusters, we realize various $d$-wave altermagnetic configurations on these non-periodic lattices. The spin-resolved spectral density difference is calculated to reveal the $d$-wave nature of the underlying magnetic orders. We then demonstrate that $d_{x^2-y^2}$-wave AM robustly induces HOTPs in these quasicrystals, with corner-state signatures that remain invariant to different types of quasicrystals. Our work not only expands the scope of AMs in quasicrystals but also establishes that the higher-order topological response may serve as a robust and intuitive tool for identifying and characterizing altermagnetic orders in non-periodic systems.

Finally, we want to emphasize that, while quasicrystals provide a particularly natural setting for demonstrating higher-order topology induced by $C_n\mathcal{T}$-symmetric altermagnetism in the absence of translational symmetry, this mechanism is not restricted to quasicrystals. In principle, it can be generalized to generic aperiodic systems\cite{zqnk-153f}, provided that the corresponding $C_n\mathcal{T}$ symmetry is preserved, as required for altermagnetism.

\begin{acknowledgments}
X.H. and H.G. acknowledges support from the NSFC grant No.~ 12574249 and the BNLCMP open research fund under Grant No. 2024BNLCMPKF023. X.Z. acknowledges support from the Natural Science Foundation of Jiangsu Province under Grant BK20230907 and the NSFC grant No.~12304177. S.F. is supported by the National Key Research and Development Program of China under Grant Nos. 2023YFA1406500 and 2021YFA1401803, and NSFC under Grant No. 12274036.
\end{acknowledgments}


\appendix
\setcounter{equation}{0}
\renewcommand\theequation{A.\arabic{equation}}
\setcounter{figure}{0}
\renewcommand{\thefigure}{A\arabic{figure}}

\bibliography{ref}

\end{document}